\documentclass[12pt]{iopart}
\usepackage{times}
\usepackage{graphicx}
\usepackage[colorlinks=true,linkcolor=black,citecolor=black]{hyperref}
\usepackage[a4paper, total={6in, 8in}]{geometry}
\usepackage{breakcites}
\usepackage{tikz}
\usepackage{csvsimple}
\usepackage{multirow}
\usepackage{mdframed}
\usepackage{hyperref}
\usepackage{parcolumns}
\usepackage{ulem}

\begin{document}
\title[]{Compiling Quantum Circuits to Realistic Hardware Architectures using Temporal Planners}

\author{Davide Venturelli$^{1,2}$, Minh Do$^{3,4}$, Eleanor Rieffel$^1$, Jeremy Frank$^4$}

\address{$^1$ NASA Ames Research Center, Quantum Artificial Intelligence Laboratory}
\address{$^2$ USRA Research Institute for Advanced Computer Science (RIACS)}
\address{$^3$ Stinger Ghaffarian Technologies (SGT Inc.)}
\address{$^4$ NASA Ames Research Center, Planning and Scheduling Group}
\ead{davide.venturelli@nasa.gov}

\maketitle

\begin{abstract}
To run quantum algorithms on emerging gate-model quantum hardware,
quantum circuits must be compiled to take into account constraints
on the hardware. For near-term hardware,
with only limited means to mitigate decoherence, it is critical to
minimize the duration of the circuit.
We investigate the application of temporal planners to the problem of compiling
quantum circuits to newly emerging quantum hardware. While our approach is
general, we focus on compiling to superconducting hardware architectures
with nearest neighbor constraints. Our initial experiments focus
on compiling Quantum Alternating Operator Ansatz

(QAOA) circuits whose
high number of commuting gates allow great flexibility in the order in
which the gates can be applied. That freedom makes it more challenging
to find optimal compilations but also means there is a greater potential win
from more optimized compilation than for less flexible circuits.
We map this quantum circuit compilation problem to a temporal planning
problem, and generated a test suite of compilation problems for QAOA circuits
of various sizes to a realistic hardware architecture.
We report compilation results from several state-of-the-art
temporal planners on this test set.
This early empirical evaluation demonstrates that temporal planning
is a viable approach to quantum circuit compilation.
\end{abstract}

\normalem

\section{Introduction}
\label{sec:intro}
We explore the use of temporal planners to optimize compilation of quantum circuits to newly emerging quantum hardware. Currently only special purpose quantum hardware is commercially available: quantum annealers that  run only one type of quantum optimization algorithm. The emerging gate-model processors, currently in prototype phase, are universal in that, once scaled up, they can run any quantum algorithm.  This facilitates expanding the empirical exploration of quantum algorithms beyond optimization, as well as enabling the exploration of a broader array of quantum approaches to optimization.

Quantum algorithms are usually specified as idealized quantum circuits that do not take into account hardware constraints.
This approach makes sense since the actual physical constraints vary from architecture to architecture.
With the advent of gate-model processors, researchers have begun to explore approached to compiling idealized quantum circuits to realistic hardware.
For example, emerging superconducting quantum processors have planar architectures with nearest-neighbor restrictions on the
locations (qubits) to which the gates can be applied. Such processors include the 5-qubit processor IBM recently made publicly available
the cloud~\cite{IBM}, recently updated to 20 qubits, and processors being fabricated by other groups, such as Intel/TU
Delft~\cite{versluis2016scalable}, UC Berkeley~\cite{o2017design}, Rigetti Computing~\cite{Sete16}  \cite{reagor2017demonstration}, and
Google~\cite{Boixo16}. All cited groups have announced plans to build gate-model quantum processors with 40 or more qubits in the near term.
Idealized circuits generally do not respect nearest neighbor constraints; idealized quantum circuits generaly contain many gates between pairs of qubits that are not nearest neighbors and therefor cannot be implemented directly on the processors. - see Figure 2 for more details.
For this reason, compiling idealized quantum circuits to superconducting hardware requires adding supplementary gates that move qubit states to locations where the desired gate can act on them.

Quantum computational hardware suffers from decoherence, which degrades the
performance of quantum algorithms over time. Especially for near-term hardware,
with only limited means to mitigate decoherence, it is critical to
\emph{minimize} the \emph{duration of (execution of) the circuit} that carries out the quantum
computation, so as to minimize the decoherence experienced by the computation.
Other, more sophisticated, compilation cost functions, such as figure-of-merits taking into 
account fidelity of operations ~\cite{onlinelink2qvolume} \cite{newIBM}, could be used in the
future within the temporal planning approach to compilation we explore here.
Optimizing the duration of compiled circuits is a challenging problem due to
the parallel execution of gates with different durations.
For quantum circuits with flexibility in when the gates can be
applied, or when some gates commute with each other (so can be applied in
a different order while still achieving the same computation) the search
space for feasible compilations is larger than for less flexible circuits.
That freedom makes it more challenging to find optimal compilations but also means there is a greater potential win
from more optimized compilation than for less flexible circuits.

While there has been active development of software libraries
 that server as a software toolchain to compile an algorithm (specified in some standardized language) into idealized quantum circuits
(see Ref.~\cite{chong2017programming} for a review, and \cite{Wecker14}
 \cite{Smith16} \cite{Steiger16} \cite{DeVitt16} \cite{Barends16} for the most relevant works),
 few approaches have been explored for compiling idealized quantum circuits to realistic quantum hardware \cite{Beals13}
\cite{brierley2015efficient}
\cite{bremner2016achieving},
leaving the problem open for innovation. 
Recent studies explore exact schemes \cite{wille2014exact}, approximate
tailored methods \cite{kole2017new} or formulations suited for off-the-shelf
Mixed Integer Linear Programming (MILP) solvers such as
Gurobi~\cite{bhattacharjee2017depth}.  The benchmarks in prior
work have mostly revolved around circuits composed by elementary gates relevant
for fault-tolerant quantum computing schemes, with attention shifting
recently in the quantum computing community towards algorithms to
be run on near-term hardware. These algorithms have circuits that often contain
a large number of mutually
commuting gates, but not necessarily natively available in the hardware.
Recently a tailored scheduling heuristic approach has been published by \cite{guerreschi2017gate} to schedule quantum circuits with many commuting gates (but only for gates of unity duration and/or on a linear architecture.)
Prior work had not used a temporal planning approach, which can be applied quite generally, enabling us to address, for the first time, gates with variable durations, generic architectures with inhomogeneous spatial features, and efficiencies that can be gained when large numbers of gates commute.  An similar issue arising when compiling classical programs is the {\em register allocation} problem, in which program variables are assigned to machine registers to improve execution time; this problem reduces to graph coloring \cite{registeralloc}.

In this paper, we apply \emph{temporal planning} techniques to the problem of
compiling quantum circuits to realistic gate-model quantum hardware.
Temporal planning is a subdomain of Automated Planning and
Scheduling, a branch of Artificial Intelligence (AI) which is concerned with the
identification of strategic decisions, among a large finite set of
possibilities, to achieve a specific goal that includes temporal constraints or
time optimization objectives.

As we will explain in Section~\ref{sec:planning_model}, we use domain-independent
AI planners to find a parallel sequence of
conflict-free instructions that, when executed, achieve the same result as the machine-independet quantum circuit.  Additional instructions
specific to a gate-model architecture are inserted.
The temporal planners aim to provide a machine-dependent plan that minimizes
makespan, while respecting all
machine-dependent constraints (e.g. available gates to perform swap operations, physical layout of the gates, duration of gates, exclusions on
simultaneous operations.)
While our approach is general, we focus
our initial experiments on circuits that have few ordering constraints and thus
allow highly parallel plans. We report on experiments using a diverse set of
temporal planners to compile circuits of various sizes to an architecture
inpired by those currently being built. This early empirical evaluation
demonstrates that temporal planning is a viable approach to quantum circuit
compilation.

In Sec.~\ref{sec:compProblem}, we describe the problem of compiling
idealized quantum circuits to specific hardware architectures in detail.
Sec.~\ref{sec:targeted-problem} describes QAOA circuits, the class of
circuits with many commuting gates that we target for our initial
invstigation.
Section \ref{sec:planning_model} explains our mapping of the quantum
circuit compilation problem to temporal planning problem.
Sec.~\ref{sec:evaluation} presents our results applying state-of-the-art
temporal planners to this circuit compilation problem.
In Sec.~\ref{sec:conclusion}, we outline future research directions
stemming from this study. 
The aim is to write the paper to communicate clearly to both the quantum computing community and the temporal planning and artificial intelligence communities, so it will necessarily contain some material that is review for one or the other group.

\section{Architecture-specific quantum circuit compilation problem}
\label{sec:compProblem}

Quantum circuits for general quantum algorithms are often described on an
idealized architecture in which any 2-qubit gate can act on any pair of qubits.
Physical constraints impose restrictions on which pairs of qubits support gate
interactions in an actual physical architecture. 
In this work, we concentrate on superconducting qubit
architectures, using an abstract model that takes into account gate durations and the nearest neighbor topology of the quantum processors which will be used as the basis for the temporal planning formulation   of the compilation problem.

In general, the model will specify different
types of abstract quantum gates, each taking different durations, with a duration depending on the
specific physical implementation in terms of \emph{primitive} gates.
Gate-model quantum architectures operate as digital computers with a clock, and times could be expressed in terms of clock cycles. Elementary or primitive gates are the gates that have been directly implemlented in the hardware and carefully calibrated, so are generally
the fastest possible operations. 
However, from a sequence of primitive gates, composite gates can be synthesized, making it possible to describe 
a chip in terms of the \emph{relevant} gates for the algorithm, as long as we take into account the number of clock cycles that are required to perform the wanted gate.

In the model,
qubits in a quantum processor can be thought of as nodes in a
graph, and the relevant 2-qubit quantum gates are associated to edges. In many
architectures more than one 2-qubit gate may be implementable between a given
pair of qubits, so this structure is a multigraph
(multiple edges allowed between two nodes).
In general the gates are non-symmetric so the multi-graph could be directed (i.e. distinguishing the roles of two qubits), as for instance is the case if we include the primitive (calibrated) gates in the IBM Quantum Experience chip~\cite{onlinelink3ibm}.

Gates that operate on distinct sets of qubits may be able operate concurrently,
though there can be additional restrictions on which operations can be done
in parallel, such as requiring the sets in operation to be non-adjacent, due to cross-talk and frequency-crowding, as in Google's
proposed architecture~\cite{Boixo16}). \\

\begin{figure}[tb]
  \includegraphics[width=\columnwidth]{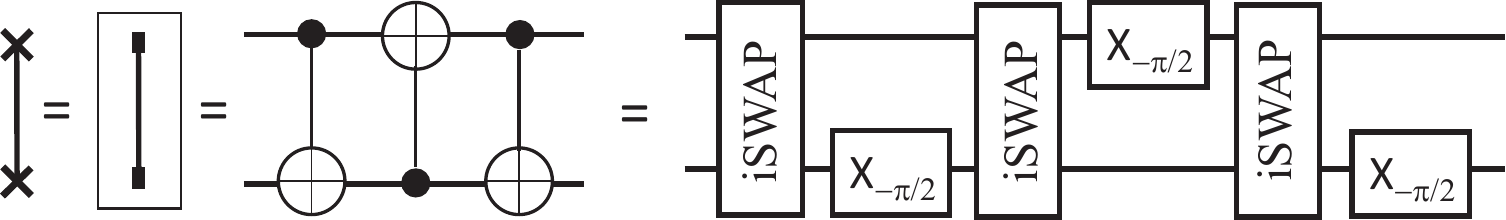}
  \caption{A known decomposition of the swap gate using as primitive gates three \emph{C-NOT}s or replacing recursively the \emph{C-NOT}s with their synthesis in terms of \emph{iSWAP}s and X-rotations.}
\label{fig:swapsynthesis}
\end{figure}

\noindent{\bf The $\textit{swap}$ gate and its synthesis:}
In this study, we make extensive use of 
a particular type of a 2-qubit gate,
the {\textit{swap}} gate, which exchanges the state of two qubits, though other gate choices are possible.
In order for the computation specified by the idealized circuit to be completed, quantum information must be moved to locations where the desired gates can be carried out, and a sequence of swap gates can be used to move the
contents of two distant qubits to a location where a desired gate can be executed. For this reason, the \emph{swap} gate is a useful gate for compilation in sparsely-connected architectures.

 This origin of the ``duration" abstraction for gates is exemplified in Figure ~\ref{fig:swapsynthesis}(left) for the \emph{swap} gate, which can be efficiently decomposed in three control-NOT ($\textit{CNOT}$) gates ~\cite{schuch2003natural}. 
The \emph{swap} gate should last at least three times as long as the $\textsc{CNOT}$ {gate~\cite{PhysRevA.69.032315}. 
Figure ~\ref{fig:swapsynthesis}(right) shows a possible synthesis of the same \emph{swap} in terms of only \emph{iSWAP} gates, which are the primitives available everywhere across the chip described in~\cite{Sete16}
. \\

Beside the timings dictated by logical gate synthesis, different choices of synchronization and time-scales of executions are possible, leading to different possible durations. For instance, in \cite{caldwell2017parametrically} the controlled-Z gate (\emph{CZ}) could last 175 or 270 nanoseconds depending on which choices are made at calibration. 
Across the chip, calibration might result in different gate times depending on the location in the chip even if the underlying circuitry is the same, as shown in~\cite{onlinelink3ibm}, where the \emph{CNOT} gate duration could vary up to a factor of 2.\\

For the purposes of
this study, we consider a simplified model in which swap gates are available between
any two adjacent qubits on the chip and all swap gates have the same duration,
but our temporal planning approach can handle the more general cases.

\subsection{Formal problem statement}

An {\it idealized quantum circuit} consists of a set of nodes (qubits), which
can be thought of as memory locations, and a specification of start times for
operations (gates), each acting on a single node or set of nodes. The
idealized quantum circuit also includes implicit or explicit specification
of which operations commute (the order in which they are executed can be
switched without affecting the computation) either individually or as blocks.
A {\it hardware architecture specification} can be viewed as a weighted,
labeled multigraph on a set of nodes, corresponding to physical quantum
memory locations (qubits) in the hardware, where each edge represents an
operation (quantum gate) that can be physically implemented on
the pair of qubits in the physical hardware, possibly as a composite gate, with the labels indicating
the type of quantum gate and the weight giving its duration. The output of the compilation process is a circuit that can be used to perform a quantum computation. It does not perform the computation itself, and therefore the compilation step can be carried out on a conventional (non-quantum) computer.
In this work, we are concerned with the efficiency and effectiveness of the compilation. A separate issue, which we do not consider here, is the performance of the quantum algorithms we are compiling.
\\

\noindent{\bf Ideal to hardware-specific quantum circuit compilation problem:}
The problem input is an idealized quantum circuit and a hardware multigraph.
The output is a time-resolved hardware-specific circuit that implements the quantum
computation described by idealized quantum circuit. The objective is to
minimize the makespan (the circuit duration) of the resulting schedule.

\subsection{Compilation examples}

Fig.~\ref{fig:hardware} shows a
hypothetical chip design that we will use for our experiments on circuit
compilation. It is inspired by the architecture proposed by
Rigetti Computing Inc.~\cite{Sete16}.  Qubits are labeled with $n_i$ and the
colored edges indicate the types of 2-qubit gates available (considering just those relevant for the algorithm), in this case swap
gates and two other types of 2-qubit gate (further described in Section 4).
Given an idealized circuit consisting only of the non-swap gates, used to
define general quantum algorithms, the circuit compilation problem is to find a
new architecture-specific circuit by adding swap gates when required,
and reordering commuting operations when desired.
The objective is to minimize the overall duration to execute all gates in
the new circuit.
To illustrate the challenges of finding effective compilation, we present some
concrete examples, with reference to the 8-qubit section in the top left of
Fig.~\ref{fig:hardware}.\\

\begin{figure}[tb]
  \includegraphics[width=\columnwidth]{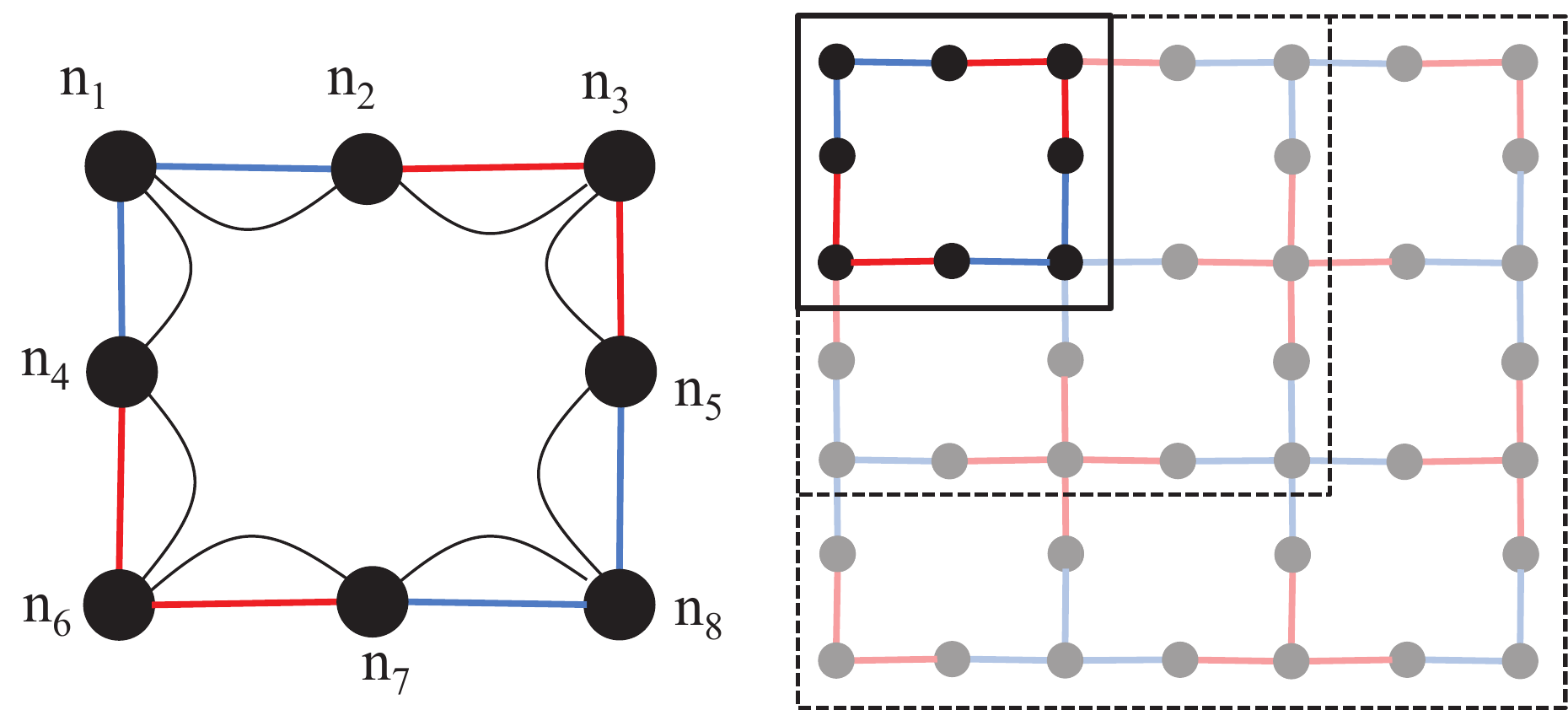}
  \caption{Left: A schematic for the hypothetical chip design,
based on an architecture proposed by Rigetti Computing 
used in our numerical experiments. Available relevant 2-qubit gates are represented
by colored arcs in a weighted multigraph. Each color is associated to
a specified, distinct gate-type and duration: \textsc{swap} gates (black) and
two other types of 2-qubits gates (red and blue). The 1-qubit gates
are present at each qubit (black dot).
Right: Dashed boxes indicate the three different chip sizes used in our
empirical evaluation (see Sec.~\ref{sec:evaluation}). For visual clarity, only
the label locations and the \textsc{swap}-gates for the smaller chip size,
corresponding to the top-left sector of the largest chip, are shown.}
\label{fig:hardware}
\end{figure}

{\bf Example 1:} Suppose that at the beginning of the compilation, each
qubit location $n_i$ is associated to the qubit state $q_i$.  Let us also
assume that the idealized circuit requires the application of a red gate to the
states $q_2$ and $q_4$, initially located on qubits $n_2$ and $n_4$.
One way to achieve this task would be to swap the state in $n_4$ with $n_1$, while at the same time swapping $n_2$ with $n_3$. Another swap, between $n_1$ and $n_2$, positions $q_4$ in $n_2$ where a red-gate connects it to $q_2$ (which is now in $n_3$).

The sequence of gates to achieve the stated goal are:
\begin{eqnarray}
&&\{\textsc{swap}_{n_4,n_1}, \textsc{swap}_{n_2,n_3}\}\rightarrow\textsc{swap}_{n_1,n_2}\rightarrow
\textsc{red}_{n_2,n_3}\nonumber\\
&&\equiv \textsc{red}(q_2,q_4)
\label{examplesequence}
\end{eqnarray}
The first line refers to the sequence of gate applications,
while the second corresponds to the algorithm objective specification (a task defined over the qubit states).
The sequence in Eq.~(\ref{examplesequence}) takes $2\tau_{swap}$ + $\tau_{red}$ clock cycles where $\tau_\star$ represents the duration of the $\star$-gate. \\

{\bf Example 2:}

Consider an idealized circuit that requires $\textsc{blue}(q_1,q_2)
\wedge \textsc{red}(q_4,q_2)$, in no particular order. If $\tau_{blue}>3 \times
\tau_{swap}$, the compiler might want to execute $\textsc{blue}_{n_1,n_2}$
while the qubit state $q_4$ is swapped all the way clockwise in five
$\textsc{swap}$s from $n_4$ to $n_3$ where $\textsc{red}_{n_2,n_3}$ can be
executed.  However, if $\tau_{blue} < 3 \times \tau_{swap}$, it is preferable
to wait until the end of $\textsc{blue}_{n_1,n_2}$ and then start to executthe instruction sequence in Eq.~(\ref{examplesequence}).\\

\section{Compiling QAOA for the MaxCut problem}
\label{sec:targeted-problem}

While our approach can be used to compile arbitrary quantum circuits to a wide
range of architectures, in this paper we concentrate on one particular case: the class of 
Quantum Alternating Operator Ansatz
(QAOA) circuits \cite{Farhi14,QAOAstuart} for MaxCut (defined in the later part of this
section) to the above-mentioned architecture inspired to 
Rigetti
Computing Inc.~\cite{Sete16}.
We choose to work
with QAOA circuits because they have many gates that commute with each other
(i.e., no ordering enforced). Such flexibility in the ordering of the gates
means that the compilation search space is larger than for other less flexible
circuits. This makes finding the optimal compilation more challenging, but
also means
there is potential for greater compilation optimization, compared to
other less flexible classes of circuits.

QAOA circuits have been the focus of recent research \cite{Farhi14} \cite{Farhi14b} \cite{Farhi16} \cite{Wecker16} \cite{Yang16} \cite{Guerr17} \cite{Jiang17}  \cite{Wang17} \cite{QAOAstuart}
in the quantum computing community since their introduction by Farhi {\it et
al.}~in \cite{Farhi14}.  The acronym was reworked from ``quantum approximate optimization algorithm" to ``quantum alternating operator ansatz" in \cite{QAOAstuart} since QAOA circuits have been applied to exact optimization and sampling as well as to approximate optimization and there are other quantum approaches to approximate optimization.

Recently Google Inc.
proposed an alternative quantum approximate optimization approach

in fixed nearest-neighbor architectures explicitly to avoid the compilation
step that is the subject of our work \cite{farhi2017quantum}.
Their numerical results on MaxCut instances show a small hit in performance.
Furthermore, unlike the QAOA circuits we consider here, in which the
number of parameters is independent of the number of qubits, their alternative
approach has many more parameters, which increase with the number of qubits,
and these parameters must be optimized separately for each architecture.
While their approach makes good use of near-term hardware with numbers of
qubits for which the parameter optimization is tractable, ultimately
one wants a scalable algorithm that can be compiled to arbitrary
architectures. Thus, while interesting, especially in the very near-term,
rather than obviating the need,
their work serves to underscore the need for efficient approaches
to optimize compilation.

We chose MaxCut as the target problem of reference, as it is becoming
one of the de facto benchmark standards for quantum optimization of all types
and it is considered a primary target for experimentation in the architecture
of ~\cite{Sete16}. \\

\begin{figure}[tb]
  \centering
  \includegraphics[width=\columnwidth]{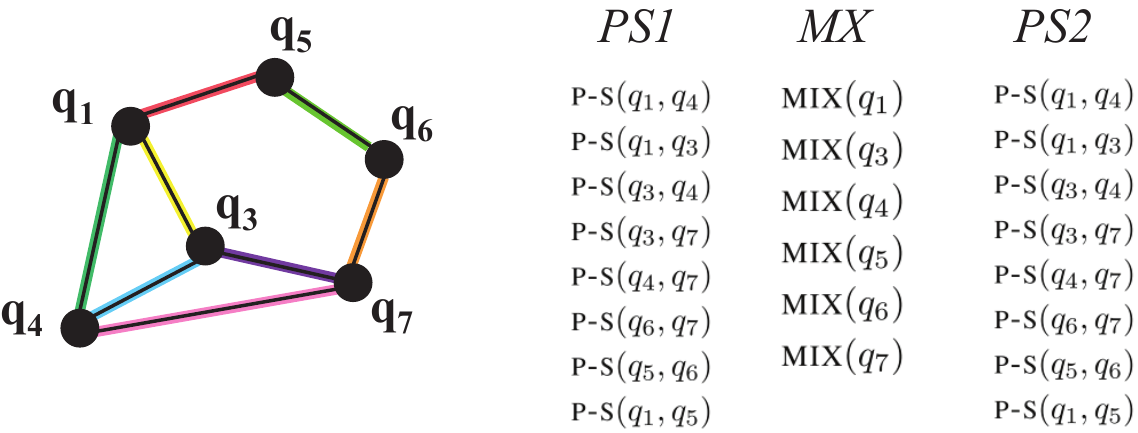}
  \caption{Example of a 6-vertex MaxCut problem on a randomly generated graph (qstates $q_2$ and $q_8$ are not appearing in this instance). 
  The association of quantum states to every node allows the definition of the compilation objectives in terms of gates, as exemplified on the right panel for QAOA $p$ = 2. Colored edges refer to Figure \ref{fig:gantt}.
  }
    \label{fig:maxcut}
\end{figure}

\noindent{\bf MaxCut Problem: }
Given a graph $G(V,E)$ with $n=|V|$ vertices and $m=|E|$ edges. The
objective is to partition the graph vertices into two sets such that the
number of edges connecting vertices in different sets is maximized.

A quadratic boolean objective function for MaxCut is:
\begin{equation}
U=\frac{1}{2}\sum_{(i,j)\in E}({1-s_i s_j}),\label{eq:QUBO}
\end{equation}
where $s_i$ are binary variables, one for each vertex $v_i$, with values
+$1$ or -$1$ indicating to which partition the vertex $v_i$ is assigned.
\\

\noindent{\bf Idealized QAOA circuits} alternate between a {\textit{phase
separation}} step (PS), based on the objective function,  and
a {\textit{mixing}} step. The phase-separation step
for QAOA for MaxCut is simpler than for other optimization problems, consisting
of a set of identical $2$-qubit gates that must be applied between certain
pairs of qubits depending on the graph of the MaxCut instance under
consideration. Specifically, the idealized QAOA circuit for MaxCut
requires a 2-qubit gate for each quadratic term in the objective function of
Eq.~(\ref{eq:QUBO}), as well as 1-qubit gates for each vertex for
the mixing step~\cite{Farhi14}.

In Fig.~\ref{fig:maxcut} a 6-vertex graph is shown, providing an
illustrative instance that will be used to describe the compilation
procedure.
We will refer to these as \emph{p-s} gates, and the main goal of
the compilation is to carry them out.

The p-s gates all commute with each
other, implying that they can be carried out in any order without changing the computation.

In the mixing phase, a set of $1$-qubit operations are applied, one to each
qubit\footnote{This is another simple feature of MaxCut, and it is
due to the fact that all possible $2^N$ $s_i$ variable assignments (see Eq.~\ref{eq:QUBO}) are defining
a valid cut. If this wasn't the case, then the mixing phase would also likely
require the application of 2-qubit gates, further complicating the scheduling
problem.}

All p-s gates that involve a specific qubit $q$
must be carried out before the mixing operator on $q$ can be applied.  These
two steps are repeated $p$ times. We consider $p$ = $1$ and $p$ = $2$ in our
experiments (detailed in Section~\ref{sec:evaluation}).

For every vertex $i\in V$, QAOA for MaxCut requires a quantum state $q_i$ to be
assigned on a qubit on the chip, and for every edge $(i,j)\in E$, the PS step
of QAOA requires executing a gate corresponding to  $\textsc{p-s}(q_i,q_j)$.
We ignore the final mixing step since it is trivial to compile by just applying
the 1-qubit mixing gate to each qubit as the last operation.

We chose the architecture proposed by Rigetti Compuing in~\cite{Sete16}
(see Fig.~\ref{fig:hardware}) for our initial exploration
because it offers a particularly interesting compilation, and
therefore planning, problem, due to the existence of two different
kinds of nearest neighbor relation in the proposed hardware.
After the synthesis of the QAOA MaxCut
gates, these two different relations become two
different durations of two-qubit gates, which corresponds to the red and blue
edges as described above.

In our problem specification, while there are
two flavors of \emph{p-s} gates (red, blue), corresponding to two different durations
of execution, the compilation goals (see figure~\ref{fig:maxcut}) do not
care on which of these two types of gates carries out the required steps.
For the purpose of this proof-of-concept work, these
durations we assign to the gates are not derived from actual designs of ongoing
experiments, but are realistic and serve to illustrate possible future designs.

The constraints on the compilation problem can be understood, with reference to Fig.~\ref{fig:hardware}, as:
\begin{itemize}
\item \textsc{swap} gates are located at every edge with $\tau_{swap}$ = 2.
\item there are two kind of non-swap gates: \textsc{p-s} gates are 2-qubit gates and \textsc{mix} gates are 1-qubit gates.
\item \textsc{p-s} gates are located at every edge of the grid, but their duration $\tau_{p-s}$ can be $3$ or $4$ depending on their location (respectively blue or red edges in Fig.\ref{fig:hardware}).
\item \textsc{mix} gates are located at every vertex with $\tau_{mix}$=1.
\item In an initialization stage, which is not considered as part of the compilation problem, a quantum state is assigned to each qubit.
\end{itemize}

\section{Compilation of a Quantum Circuit as Temporal Planning Problem}
\label{sec:planning_model}

Planning is the problem of finding a conflict-free set of actions and their respective execution times that connects the \emph{initial-state} $I$ and the desired \emph{goal state} $G$. We now introduce some key concepts that provide the background for approaching the problem of compiling quantum circuits as a temporal planning problem.

Classical planning problems are expressed in terms of binary {\it state
variables\/} and {\it actions}.  Examples of state variables for our problem
are ``The quantum state $\Psi$ is assigned to qubit number $X$" and ``The
quantum state $\Phi$ has been transformed by the application of gate $G$
present on qubits $X$ and $Y$," which may be True or False.
Actions consist of two lists, a
set of {\it preconditions} and a set of {\it effects}.

The effects of an action consists of a
subset of state variables with the values they take on if the action is carried
out. For example, the action ``State $\Psi$ is now moved from qubit $X$ to qubit
$Y$" has one precondition, ``State $\Psi$ is assigned to $X$ = True" and has two
effects ``State $\Psi$ is assigned to $X$ = False" and ``State $\Psi$ is assigned to
$Y$ = True."

A specific planning problem specifies an {\it initial state\/}, with values
specified for all state variables, and a {\it goal\/}, specified values for one
or more state variables. As for preconditions, goals are conventionally
positive, so the goal value for the goal variables is True. Generally, the
goal specifies values for only a small subset of the state variables. A plan is
a sequence of actions.

A valid plan, or a solution to the planning problem, is
a sequence of actions $A_1, ..., A_L$ such that the state at time step
$t_{i-1}$ meets the preconditions for action $A_i$, the effects of action $A_i$
are reflected in the state at time step $t_i$, and the state at the end has all
of the goal variables set to True. For an introduction on Automated Planning
and Scheduling, see \cite{planningbook}.\\

\noindent{\emph{Planners:}
A planner is software implementing a collection of algorithms; it takes as
input a specification of domain and a problem description and returns a
valid plan if one exists. Many different approaches have been implemented to
find a viable plan, among them: (i) heuristically search over the possible
valid plan trajectories or over the library
of partial plans or (ii) compile the planning problem into another combinatorial substrate (e.g., SAT, MILP, CSP) and feed the problem to off-the-shelf solvers.\\

\noindent \emph{Planning Domain Description Language (PDDL):} PDDL is a
modeling language that was originally created to standardize the input for
planners competing in the International Planning Competition (IPC). Over time,
it has become the de facto standard for modeling languages used by many
domain-independent planners. We use PDDL 2.1, which allows the modeling of
temporal planning formulation in which every action $a$ has duration $d_a$,
starting time $s_a$, and end time $e_a = s_a + d_a$.  Action conditions
\emph{cond}(a) are required to be satisfied either (i) instantaneously at $s_a$
or $e_a$ or (ii) required to be true starting at $s_a$ and remain true until
$e_a$. Action effects \emph{eff}(a) may instantaneously occur at either $s_a$
or $e_a$. Actions can execute when their temporally-constrained conditions are
satisfied, and when executed, will cause state-change effects.  The most common
objective function in temporal planning is to minimize the plan
\emph{makespan}, i.e. the shortest total plan execution time. This objective
matches well with the objective of our targeted quantum circuit compilation
problem.  To enable reuse of key problem features present in an ensemble of
similar instances, the PDDL model of a planning problem is separated into two
major parts: (i) the \emph{domain} description that captures the common objects
and behaviors shared by all problem instances of this planning domain and (ii)
the \emph{problem instance} description that captures the problem-specific
objects, initial state, and goal setting for each particular problem.\\

\begin{figure}[htbp]
  \centering
  \begin{minipage}[t]{0.5\textwidth}
  \footnotesize{
  (:constants  q1 q2 q3 q4 q5 q6 q7 q8  - qstate)\\
  
  (:durative-action swap\_1\_2\\
   \hspace*{0.5cm}:parameters (?q1 - qstate ?q2 - qstate)\\
   \hspace*{0.5cm}:duration (= ?duration 2)\\
   \hspace*{0.5cm}:condition  (and  (at start (located\_at\_1 ?q1))\\
    \hspace*{2.4cm}       (at start (located\_at\_2 ?q2)))\\
   \hspace*{0.5cm}:effect   (and (at start (not (located\_at\_1 ?q1)))\\
    \hspace*{1.95cm}        (at start (not (located\_at\_2 ?q2)))\\
    \hspace*{1.95cm}        (at end (located\_at\_1 ?q2))\\
    \hspace*{1.95cm}        (at end (located\_at\_2 ?q1))))\\ \\

(:durative-action P-S\_1stPhaseSeparation\_at\_6-7 \\
\hspace*{0.5cm}   :parameters (?q1 - qstate ?q2 - qstate) \\
\hspace*{0.5cm}   :duration (= ?duration 3) \\
\hspace*{0.5cm}   :condition    (and (at start (located\_at\_6 ?q1)) \\
\hspace*{2.4cm}        (at start (located\_at\_7 ?q2)) \\
\hspace*{2.4cm}        (at start (not (GOAL\_PS1 ?q1 ?q2))) 

\hspace*{0.5cm}   :effect   (and (at start (not (located\_at\_6 ?q1))) \\
\hspace*{1.95cm}           (at start (not (located\_at\_7 ?q2))) \\
\hspace*{1.95cm}           (at end (located\_at\_6 ?q1)) \\
\hspace*{1.95cm}           (at end (located\_at\_7 ?q2)) \\
\hspace*{1.95cm}           (at end (GOAL\_PS1 ?q1 ?q2)) \\
\hspace*{1.95cm}           (at end (GOAL\_PS1 ?q2 ?q1)))))\\

   }
 \end{minipage}%
\begin{minipage}[t]{0.5\textwidth}
\footnotesize{

(:durative-action mix\_q5\_at\_1\\
   \hspace*{0.5cm}:parameters ( )\\
   \hspace*{0.5cm}:duration (= ?duration 1)\\
   \hspace*{0.5cm}:condition (and (at start (located\_at\_1 q5))\\
   \hspace*{2.4cm}     (at start (GOAL\_PS1 q1 q5))\\
   \hspace*{2.4cm}     (at start (GOAL\_PS1 q5 q6))\\
   \hspace*{2.4cm}     (over all (not (mixed q5))))\\
   \hspace*{0.5cm}:effect (and (at start (not (located\_at\_1 q5)))\\
   \hspace*{1.95cm} (at end (located\_at\_1 q5))\\
   \hspace*{1.95cm} (at end (mixed q5))))\\

(:durative-action P-S\_2ndPhaseSeparation\_at\_6-7 \\
\hspace*{0.5cm}   :parameters (?q1 - qstate ?q2 - qstate) \\
\hspace*{0.5cm}   :duration (= ?duration 3) \\
\hspace*{0.5cm}   :condition    (and (at start (located\_at\_6 ?q1)) \\
\hspace*{2.4cm}        (at start (located\_at\_7 ?q2)) \\
\hspace*{2.4cm}        (at start (not (GOAL\_PS2 ?q1 ?q2))) \\
\hspace*{2.4cm}        (at start (GOAL\_PS1 ?q1 ?q2)) \\
\hspace*{2.4cm}        (at start (mixed ?q1)) \\
\hspace*{2.4cm}        (at start (mixed ?q2))) \\
\hspace*{0.5cm}   :effect       (and (at start (not (located\_at\_6 ?q1))) \\
\hspace*{1.95cm}           (at start (not (located\_at\_7 ?q2))) \\
\hspace*{1.95cm}           (at end (located\_at\_6 ?q1)) \\
\hspace*{1.95cm}           (at end (located\_at\_7 ?q2)) \\
\hspace*{1.95cm}           (at end (GOAL\_PS2 ?q1 ?q2)) \\
\hspace*{1.95cm}           (at end (GOAL\_PS2 ?q2 ?q1)))))
}
  \end{minipage}

  \caption{ PDDL model of actions representing some exemple of \textsc{swap}, \textsc{mix}, and \textsc{p-s} gates. The first line indicates that this compilation problem involves $8$ qubit states. For each action, the duration indicates how long the action takes. The first action, a swap at qubits $1$ and $2$, has as parameters the two qstates to swap. The condition checks that these states are indeed located at the qubits on which the swap will occur. The effect makes sure the states have been swapped. The second action mixes qstate $q5$ at qubit $1$, with conditions that state $q5$ is indeed at qubit $1$, and both the phase separation gates involving qstate $q5$ (see Fig.~\ref{fig:maxcut}) have been carried out. The effects include setting $mixed\; q5$ to \textsc{TRUE} The third and fourth actions are phase separation actions in a $p = 2$ circuit corresponding to the first and second levels of the algorithm.
}
    \label{fig:PDDL}
\end{figure}

PDDL is a
flexible language that offers multiple alternatives for modeling a planning
problem. These modeling choices greatly affect the performance of existing
PDDL planners. For instance, many planners pre-process the original domain
description before building plans; this process is time-consuming, and
may produce
large `ground' models depending on how action templates were written.  Also,
not all planners can handle all PDDL language features effectively (or even at
all).  For this project, we have iterated through different modeling choices
with the objective of constructing a PDDL model that: (i) contains a small
number of objects and predicates for compact model size; (ii) uses action
templates with few parameters to reduce preprocessing effort; while  (iii)
ensuring that the model can be handled by a wide range of existing PDDL
temporal planners.\\

\noindent{\bf Modeling Quantum Gate Compilation in PDDL 2.1:}
To apply a temporal planner to the circuit compilation problem, we must
represent the allowed gates as actions and the desired circuit as a
set of goal variables.
We describe how to do so for the QAOA circuit compilation problem
exemplified in Fig.~\ref{fig:maxcut}, ensuring that for a plan to be valid,
the required $\textsc{p-s}$ or $\textsc{mix}$ gates are scheduled for
each step of the algorithm.
At the high-level, in this domain, we need to model: (i) how
actions representing \textsc{p-s}, \textsc{swap}, and \textsc{mix} gates affect
qubits and qubit states (qstate); (ii) the actual qubits and qstates involved
in a particular compilation problem, with their initial locations and final
goal requirements, (iii) the underlying graph structure (gates connecting
different pairs of qubits). We follow the conventional practice of modeling (i)
in the domain description while (ii) is captured in the problem description.
One common practice is to model (iii) within the problem file. However, given
that we target a rather sparse underlying qubit-connecting graph structure (see
Fig.~\ref{fig:hardware}), we decide to capture it within the domain file to
ease the burden of the ``grounding'' and pre-processing step for existing
planners, which can be very time-consuming. Specifically:\\

\noindent{\emph{Objects:}} We need to model three types of object: qubits,
qstates, and the location of the \textsc{p-s} and \textsc{swap} gates (i.e.,
edges in the multigraph of Fig.~\ref{fig:hardware}
connecting different qubits). Since qstates are associated (by means of
the predicate $located\_at$}, see Fig.~\ref{fig:PDDL} for concrete example) to
specific qubits, they have been modeled explicitly as planning objects, while
the qubits and the gate locations (i.e., edges) are modeled implicitly.  It is
clear from the action definitions in Fig.~\ref{fig:PDDL} that qubit locations
are embedded explicitly within the action declaration.  This approach avoids
declaring qubits as part of the action parameters,
significantly reducing the number of ground actions to be generated. For
2-qubit actions, the potential number of ground actions reduces from $N^4$ to
$N^2 \times |E|$, with $N$ the number of qubits in the chip (up to 40) and $E$ the set of connections between qubits. While it's true that many modern
planners will be able to filter out invalid ground actions during the
grounding/preprocessing step, our empirical evaluation shows that capturing the graph structure explicitly in the domain file speeds up the preprocessing time of all tested planners, sometime as significantly as 40x.\\

\noindent{\emph{Actions:}} Temporal planning actions are created to model: (i)
2-qubit \textsc{swap} gates, (ii) 2-qubit \textsc{p-s} gates, and (iii) 1-qubit
\textsc{mix} gates. For reference, Fig.~\ref{fig:PDDL} shows the PDDL
description of a \textsc{swap} gate between qubits 1 and 2, the \textsc{mix}
gate of state $q_5$ on qubit 1, and the \textsc{p-s} gates between qubits 6 and
7 at the first and second phase separation.
\footnote{The full set of PDDL model for all our tested problems is available
at:
\url{https://ti.arc.nasa.gov/m/groups/asr/planning-and-scheduling/VentCirComp17\_data.zip}}
In the action's condition list, we
specify that gates are accomplished on the two qstates only if they are located
on the corresponding qubits. Note that some planners (e.g., CPT, POPF) can not handle
action preconditions that are specified negatively (e.g., $(over all (not (mixed q5)))$ of action mix\_q5\_at\_1 in Figure~\ref{fig:PDDL}). For those planners, we've also created another PDDL version of our domain
where dummy predicates such as $not\_mixed$ is introduced to represent the opposite value of those
that need to be negatively satisfied in some action's precondition list.  To prevent a qstate $q$ currently belonging to
qubit $X$ from being addressed by multiple gates at the same time (i.e.
``mutex" relations in planning terminology), we assign value \texttt{FALSE} to
the predicate $(located\_at\_X\; q)$ at the starting time of all actions
involving $q$.

The most complex constraint to model is the  conditions to mix a qstate $q$ given the requirement that \emph{all} \textsc{p-s} gates involving $q$ in the previous phase separation step have been executed. We explored several other choices to model this requirement such as: (i) use a metric variable $PScount(q)$ to model how many \textsc{p-s} gates involving $q$ have been achieved at a given moment; or (ii) use ADL quantification and conditional effect constructs supported in PDDL.  Ultimately, we decided to explicitly model all \textsc{p-s} gates that need to be achieved as conditions of the $\textsc{mix}(q)$ action. This is due to the fact that alternative options require using more expressive features of PDDL2.1 which are not supported by many effective temporal planners.\footnote{Only one of six planners in the Temporal track of the latest IPC (2014) supports numeric variables and also only one of six supports quantified conditions. Preliminary tests with our PDDL model using metric variables to track satisfied goals involving qstate $q$ using several planners shows that they perform much worse than on non-metric version, comparatively. This is to be expected as currently, state-of-the-art PDDL planners still
 do not handle metric quantities as well as logical variables.}\\

\noindent{\emph{Objective:}} For a level $p$ QAOA circuit, the goal is to have all of the $(GOAL\_PSi\; ?q1\; ?q2)$ predicates, for any $q1$ and $q2$ connected in the graph, for $1 \leq i \leq p$ set to \textsc{TRUE} and all $mixed_i\; qj$ set to true for $1 \leq j \leq N$ and  $1 \leq i \leq p - 1$ (since the final mixing step can be added by hand at the end). Since we only consider $p = 1$ and $p = 2$, so only have a mixing step in the $p = 2$ case, we have simplified the notation to simply $mixed\; qj$. Further, we use the standard temporal planning objective of minimizing the plan \emph{makespan}. Minimizing the makespane coincides with  minimizing the \emph{circuit depth}, which is the main objective of the compilation problem.\\

\noindent{\bf Alternative models:} Given that non-temporal planners can perform
much better than temporal planners on problems of the same size, we also
created the non-temporal version of the domain by discretizing action durations
into consecutive ``time-steps'' $t_i$,  introducing additional predicates
\emph{next}$(t_i, t_{i+1})$ enforcing a link between consecutive time-steps.
However, initial evaluation of this approach with the M/Mp SAT-based
planner~\cite{rintanen:aij12} (which optimize parallel planning steps)
indicated that the performance of non-temporal planners on this discretized
(larger) model is much worse than the performance of existing temporal planners
on the original model.

Another option is to totally ignore the temporal aspect
and encode it as a ``classical'' planning problem where actions are
instantaneous. A post-processing step is then introduced to inject back the
temporal constraints and schedule actions in the found classical plans. While
we do not believe this approach would produce good quality plans, it's another
promising option to scale up to larger problems in this domain.

\section{Empirical Evaluation}
\label{sec:evaluation}

We modeled the QAOA circuit compilation problem as described in the
previous sections and tested them using various off-the-shelf PDDL 2.1 Level 4
temporal planners. The results were collected on a RedHat Linux 2.4Ghz machine
with 8GB RAM.\\

\noindent{\em Problem generation:} We consider three problem sizes based on
grids with N = 8, 21 and 40 qubits (dashed boxes in Fig.~\ref{fig:hardware}).
The utilized chip layouts are representative of devices to come in the next
2 years\footnote{A gate-model 8-qubit chip with the grid we used is currently available from Rigetti, however the gate set is currently uncalibrated or calibrated with different durations depending on the edges, so our benchmarks do not model actual hardware.}

For each grid size, we generated two problem classes: (i) $p$ = 1 (only one
PS-mixing step) and (ii) $p$ = 2 (two PS-mixing steps). To generate the graphs
$G$ for which a MaxCut needs to be found, for each grid size, we randomly
generate 100 Erd\"os-R\'enyi graphs $G$ \cite{ErRe}.   Half (50 problems) are
generated by choosing $N$ of $N(N-1)/2$ edges over respectively 7, 18, 36
qstates randomly located on the circuit of size 8, 21, and 40 qubits (referred
to herafter as `Utilization' u=90\%). The other half are generated by choosing
$N$ edges over 8, 21, and 40 qstates, respectively (referred to herafter as
`Utilization' u=100\%). In total, we report tests on 600 random planning
problems with size in the range [1024-232000] for the number of grounded actions and [192-8080] for the number of predicates.\\

\noindent{\em Planner setup:} Since larger $N$ and $p$ lead to more complex
setting with more predicates, ground actions, requiring planners to find
longer plans, the allocated cutoff time for different setting are as follow:
(i) 10 minutes per instance for $N = 8$, (ii) 30 minutes per instance 
for $P = 1, N = 21$; (iii) 60 minutes per instance  for other 
cases. The timelimits are comparable to what
used in the previous International Planning Competitions. We select planners that performed well in the
temporal planning track of previous IPCs, while at the same time representing a
diverse set of planning technologies:
\begin{itemize}
\item \emph{LPG}: which is based on local search with restarts over action
graphs \cite{LPG}. Specifically, LPG incrementally builds a
multi-level graph structure. Each layer represented by a single action
and each graph edge represents a supporting connection between one action's
effect with a condition of another action appearing in a later layer. The graph
leaf nodes represent action conditions that have not been supported (i.e.,
``connected'') by other action effects. At the beginning of the search process,
LPG starts with a two-layer graph consisting of two newly created actions: (i)
$A_{init}$: which occupies the first layer of the graph, has an empty condition
list, and has an effect list represents state variables that are true in the
initial states; (ii) $A_{goal}$: which occupies the last layer, has an empty
effect list, and has a condition list represents all goals. At each search
step, LPG generates the local search neighborhood by considering all decisions
of either: (i) establishing a new edge connecting an existing action's effect
with an open condition of another action appears in a later layer (without
conflicting with negative effects of other actions), (ii) removing an edge from
the existing graph; (iii) adding another action to the graph; (iv) removing an
action from the graph. Each resulting candidate partial (i.e., incomplete) plan
in the local search neighborhood is evaluated by a heuristic function balancing
between how close that candidate is from being a complete plan (i.e., fewer
unsatisfied conditions) and how good the quality of the likely complete plan
starting from that candidate partial plan based on the user's defined objective
function (e.g., minimizing the plan makespan). LPG then selects the best
candidate partial plan from the search neighborhood and starts a new search
episode. This process is repeated until a complete plan is found. When LPG is
run in the ``anytime'' mode, it does not stop when the first complete plan is
found, but will restart its planning process with the found plan(s) used as the
baseline quality comparison on the subsequent trials.

\item \emph{Temporal FastDownward (TFD):} a heuristic forward state-space (FSS) search planner with post-processing to reduce makespan \cite{TFD}. In the FSS framework, the planner starts from the initial state $I$ with an empty plan $P$ and tries to extends $P$ until the state resulted from applying $P$ satisfies all goals\footnote{This is in contrast to the backward state-space (BSS) planners, which build the plan ``backward'' starting from the goals until it reaches the initial state.}. In each search step, FSS planners will generate new search nodes by taking a state $S_P = Apply(P,I)$, reached from applying $P$ to the initial state $I$, and considers all actions $A$ applicable in $S_P$ (i.e., all conditions of $A$ are satisfied by $S_P$). All newly generated states $S' = Apply(A,S_P)$ are put in the search queue, ordered by the heuristically evaluated ``quality'' of $S'$. The heuristic value evaluating a given state $S$ generally depends on two factors: (i) the quality of the partial plan leading from $I$ to $S$, and (ii) the estimation on the quality of the remaining plan leading from $S$ to the goals. In TFD, the second part is estimated through analyzing a set of special structure called the domain-transition graphs (DTG) and causal-graph (CG) that are statically built for each planning problem\footnote{A directed edge in the DTG connects two values $v$ and $v'$ of a given state variable $s$ in which there exist an action $a$ that can make the ``transition'' from $v$ to $v'$ by deleting $v$ and add $v'$ when executed. There is an edge in CG connecting two DTGs associated with two state variables $s$ and $s'$ if there is an action $a$ that has a condition depends on $s$ and an effect causing change of the value of $s'$.}. After a valid plan $P$ is found, TFD also tries to improve the final plan makespan by rescheduling actions in $P$, pushing them to start as early as possible without violating the various logical and temporal constraints between different actions in $P$ such as causal supports and potential conflicts caused by actions' negative effects. This post-processing step is done greedily and takes little time compared to the planning process.

\item \emph{SGPlan:} partition the planning problem into subproblems that can be solved separately, while resolving the inconsistencies between partial plans using extended saddle-point condition \cite{SGPlan1} \cite{SGPlan2}. Specifically, SGPlan uses a sub-goal partitioning strategy in which a high-level planning problem is divided into smaller planning problems, each one targets a smaller subset of goals. Furthermore, if a ``landmark'' (i.e., a given state or condition that needs to be visited by all plans when solving a given problem) is found for a subset of the goals, that landmark can be used to further partition a sub-planning problem into a subset of secondary sub-problems. Thus, the original planning problem can be partitioned into a hierarchy of multi-level interconnected smaller sub-problems, each with its own initial state and set of goals. Each sub-problem can be solved by any off-the-shelf planner. In particular, SGPlan uses a slightly modified Metric-FF, a forward state-space planner, and an earlier version of LPG to solve sub-planning problems.

\item \emph{CPT:} uses the Partial Order Causal Link (POCL) framework, which once dominated planning research. POCL planners search through the space of {\em partial plans}; each one consists of a list of actions and the causal-link between them. A {\em causal-link} indicates that an action's effect is used to support another action's condition. CPT~\cite{vidal:aij2006} utilizes techniques from constraint-programming  to create (1) effective branching scheme to select what action to consider next during each search step; (2) a makespan-bound automatically extracted from the planning problem to set the horizon for the solution search. 

At the moment, CPT is one of the most effective temporal planners that can minimize plan makespan.

\item \emph{POPF:} is a forward-chaining planner that combines forward-state-space search framework with ideas from the POCL planning framework. In POCL planning, a new action is a {\em threat} to a causal link if it removes the condition the action introduces in support of another action. During the forward search, when applying an action to a state, POPF~\cite{popf} seeks to introduce only the ordering constraints needed to resolve threats, rather than insisting the new action occurs after all of those already in the plan. POPF mostly uses the relaxed-plan heuristic similar to other forward state-space planners, but has a dedicated Simple Temporal Network (STN) 
implementation to handle the temporal constraints incurred from temporal actions interleaving in the explored partial plans. Besides temporal constraints, POPF can also handle linear continuous numeric effects on resources. It does so by offloading those constraints to a dedicated MILP solver.
\end{itemize}

We ran SGPlan (Ver 5.22), TFD (Ver IPC2014), POPF, and CPT (latest versions at time of writing) with their default parameters. Unlike other planners, which support a single mode, LPG (Ver TD 1.0) has three standard modes, which we all ran to collect empirical results: (i) \emph{-speed} that uses heuristic geared toward finding a valid plan quickly, (ii) \emph{-quality} that uses heuristic balancing plan quality and search steps, and (iii) \emph{-n 10} ($k =10$) that will try to find within the time limit up to \emph{10} plans of gradually better quality by using the makespan of previously found plan as upper-bound when searching for a new plan. Since LPG ($k = 10$) option always dominates both LPG-quality and LPG-speed by solving more problems with better overall quality for all setting, we will exclude results for LPG-quality and LPG-speed from our evaluation discussion. For the rest of this section, LPG result is represented by LPG ($k = 10$).\\

\begin{table}[t]
\centering
{
\begin{tabular}{|l|c|c|c|c|c|c|c|c|c|c|}
\hline
\multirow{2}{*}{} & \multicolumn{6}{c|}{P1}                                                       & \multicolumn{4}{c|}{P2}                            \\ \cline{2-11}
                  & \multicolumn{2}{c|}{N8} & \multicolumn{2}{c|}{N21} & \multicolumn{2}{c|}{N40} & \multicolumn{2}{c|}{N8} & \multicolumn{2}{c|}{N21} \\ \hline
Utilization $u$   & 0.9      & 1.0      & 0.9     & 1.0    & 0.9    & 1.0    & 0.9    & 1.0      & 0.9     & 1.0  \\ 
\hline 
\hline
LPG               & 50       & 50       & 50      & 50     & 10     & 14     & 50     & 50      &    50   &    50  \\ \hline
TFD               & 50       & 50       & 50      & 50     & -      & -      & 50     & 50       & 50      & 50   \\ \hline
SGPlan            & 50       & 50       & 50      & 50     & 50     & 50     & 50     & 50       & -       & -    \\ \hline
POPF              & 50       & 50       & 48      & 50     & 8      & 19      & 50     & 50       & 4       & 6   \\ \hline
CPT               & 50       & 50       & -       & -      & -      & -      & -      & -        & -       & -   \\ \hline
\end{tabular}

\caption{Summary of the solving capability of selected planners. Numbers indicate how many random problems out of 50 have been solved.
}\label{tab:overall_result}}

\end{table}

\begin{table}[tbp]
\centering
\begin{tabular}{|l|c|c|c|c|c|c|}
\hline
                  & \multicolumn{2}{c|}{p=1, N8} & \multicolumn{2}{c|}{p=1, N21} & \multicolumn{2}{c|}{p=2, N8} \\ \hline
Utilization $u$   & 0.9       & 1.0       & 0.9        & 1.0       & 0.9      & 1.0        \\ 
\hline 
\hline
LPG               &{\bf 0.88} &{\bf 0.89} & 0.91       & 0.87      & 0.50     & 0.52       \\ \hline
TFD               & 0.87      & 0.87      &{\bf 0.95}  & 0.90      &{\bf 0.99} &{\bf 0.99} \\ \hline
SGPlan            & 0.67      & 0.68      & 0.64       & 0.67      & 0.75     & 0.79       \\ \hline
POPF              & 0.81      & 0.81      & 0.90       &{\bf 0.94} & 0.87     & 0.91       \\ \hline

\end{tabular}
\caption{Plan quality comparison between different planners using IPC formula (higher value indicates better plan quality). Highlighted results represent the best performance in the problem class. CPT results for N=8 p=1 are proven optimal so they are implicitly assigned 1.0.}

\label{tab:plan_quality}
\end{table}

\begin{figure*}[tb]
  \centering
  \includegraphics[width=\columnwidth]{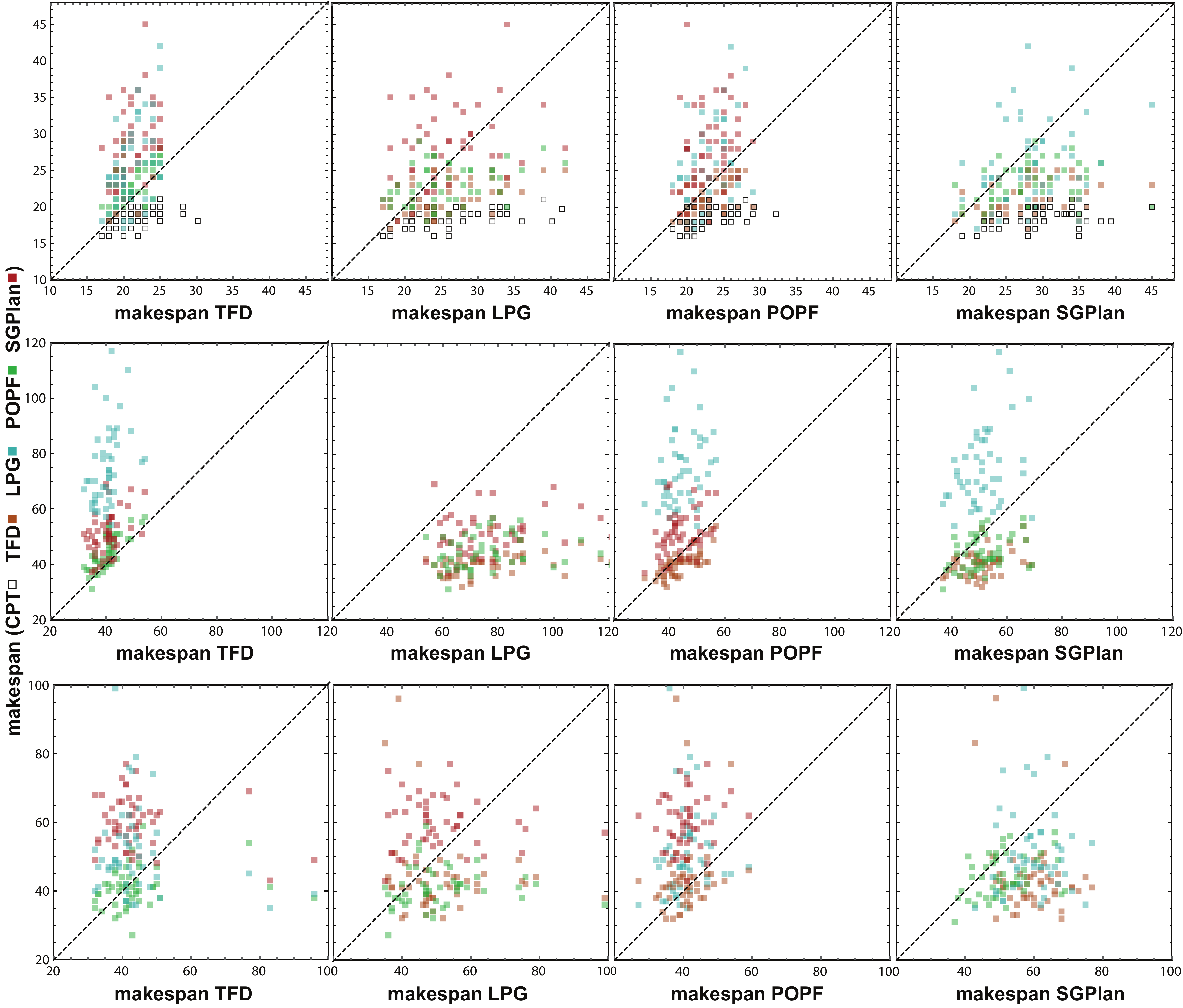}
  \caption{Instance-by-instance makespan comparison of the used planners on the problem set for $u$=1.0 (results for $u$=0.9 are qualitatively similar). Scatterplots indicate on the x-axis a specific planner, compared on the y axis against another planner (color code in the legend). The dashed line indicate equal makespan. The first row of plots shows $N$=8, $p$=1; the second row $N$=8, $p$=2; the third row $N$=21, $p$=1.}

    \label{fig:scatterplots}
\end{figure*}

\noindent{\bf Evaluation Result Summary:} Table~\ref{tab:overall_result} shows
the overall performance on the ability to find a plan of different planners.
SGPlan stops after finding one valid plan while TFD, LPG, and POPF are ``anytime'' planners that exhaust the
allocated time limit and try to find gradually improving quality plans. 
While CPT can find multiple plans, it does not return any until it can prove that the plan found is optimal. Since
no planner was able to find a single solution for $N$ = 40 and $p$ = 2
within the $60$ minute cutoff, we omit
the result for this case from Table~\ref{tab:overall_result}. Overall, LPG was able to solve the highest
number of problems, followed by TFD, SGPlan, and POPF. Being an optimal-guarantee planner, CPT can only solve
the smallest problem set ($N = 8$, $p = 1$) and can not find any solution for the other sets.
SGPlan can find a solution very quickly, compared to the time it takes other
three other anytime planners to find the first solution. It is the only planner that can scale up and solve all 100 problem} in the $N = 40$ for $p$ = 1 (finding plans with 150-220 actions). Unfortunately,
SGPlan stopped with an internal error for $N$ = 21 and  $p$ = 2. TFD generally
spent a lot of time on preprocessing for $p = 1, N = 21$  (around 15 minutes)
and $p = 2, N = 21$ (around 30 minutes) but when it is finished with the
pre-processing phase\footnote{The two most time-consuming parts in
TFD's pre-processing routine are ``processing axioms'' and ``invariant
analysis''. While ``processing axioms'' are always consistently time-consuming,
``invariant analysis'' is heuristically done and sometime can be quick while
some other times can be very time consuming.} it can find a solution
quickly and also can improve the solution quality quickly. TFD spent all
of the 60 minutes time limit on pre-processing for $N = 40$ problems. LPG
can generally find the first solution more quickly than POPF and much faster than TFD (but still much
more slowly than SGPlan) but does not improve the solution quality as quickly as TFD or POPF. \\

\noindent{\em Plan quality comparison:}  to compare the plan quality across planners, we use the formula employed by the IPCs to grade planners in the temporal planning track since  IPC6~\cite{ipc2008}: for each planning instance $i$, if the best-known makespan is produced by a plan $P_i$, then for a given planner $X$ that returns a plan $P^i_X$ for $i$, the score of $P^i_X$ is calculated as: $makespan(P_i)$ divided by $makespan(P^i_X)$. A comparative value closer to 1.0 indicates that planner $X$ produces better quality plan for instance $i$. We use this formula and average the score for our three tested planners over the instance ensembles that are completely solved by the time cutoff. Table~\ref{tab:plan_quality} shows the performance of different planners with regard to plan quality. For $N$ = 8 and $p$ = 1, for which we know the optimal plans given that CPT was able to solve all 100 problems, LPG found the best quality plans but TFD is only slightly worse and POPF is not far behind.  

The comparison results for $N$ = 21 and $p$ = 1 is similar in the sense that the three anytime planners LPG, TFD, and POPF perform very similarly but in this case TFD and POPF are slightly better than LPG. For $N$ = 8 and $p$ = 2, TFD nearly always produce the best quality plan with POPF slightly behind while LPG perform significantly worse. The mixing and the requirement to synchronize the two phase-separation seems to confuse the local search heuristic in LPG. SGPlan, which unlike TFD, LPG, POPF only find a single solution, produce lower quality plans, as expected. However, for the $p$ = 2 case, SGPlan produces overall better quality plans compared to LPG, even though LPG returns multiple plans for each instance. 

Fig.~\ref{fig:scatterplots} shows in further detail the head-to-head makespan comparison between different pairs of planners, specifically pairwise comparisons between TFD, SGPLan, LPG, and POPF: TFD always dominates SGPlan, TFD dominates LPG majority of the times (except for $N$ = 8, $p$ = 1 where the performances are comparable), and SGPlan dominates LPG on bigger problems, but is slightly worse on smaller problems. POPF performance in general is very similar with the one of TDF, especially for $N$ = 8.

For a more detailed analysis with data reported for each specific instance, see ~\cite{onlinelink}.\\

\noindent{\em Planning time comparison:}  Both TFD, LPG, and POPF use ``anytime" search algorithms and use all of their allocated time to try finding better gradually better quality plans. In contrast, SGPlan return a single solution and thus generally take a very short amount of time with the median solving time for SGPlan in $p$=1$|N_8$, $p$=1$|N_{21}$, $P$=1$|N_{40}$ and $P$=2$|N_8$ are 0.02, 1, 25, and 0.05 seconds\footnote{For comparison purpose, LPG-quality, which also try to returns a single solution of good quality, produces the median solving time for $P$=1$|N_8$ and $P$=2$|N_8$ are 0.9 and 70 seconds respectively.}. CPT, in general, set a very tight upper-bound on makespan before trying to find a plan within the bound and prove that the solution is optimal. For the smallest problem set when it can solve all 100 instances, it took a very short time between 1 to 2 seconds to find and prove optimality. However, for any other bigger set, its tight upper bounds proven to be ineffective in finding a single solution.\\

\noindent{\bf Other planners:} We have also conducted tests on: \emph{VHPOP}, \emph{HSP*}, and \emph{YASPH}. While LPG, SGPlan, TFD, and POPF were selected for their ability to solve large planning problems, we hoped that HSP* and VHPOP would return optimal plans to complement CPT in providing a baseline for plan quality estimation. Unfortunately, HSP* and VHPOP failed to find a single plan even for our smallest problems for various reason: VHPOP ran out of memory quickly, while HSP* couldn't find any plan for a cutoff time of 2 hours. We have preliminary acceptable results with YAHSP up to N=40 but they will be discussed in a future work.\\

\noindent{\bf Discussion:} Our preliminary empirical evaluation shows that the
test planners provide a range of tradeoffs between scalability and plan
quality. At one end, SGPlan can scale up to large problem sizes and
solve them in a short amount of time, providing reasonably good quality
plans (compared to the best known solutions). At the other end, TFD
utilizes all of the allocated time to find the best quality solutions but
in general is the slowest by far to obtain a valid solution. LPG and POPF balance between the two: they can either find one solution quickly like SGPlan or can utilize the whole time prior to cutoff to find better quality solutions.

Since planning is exponentially hard with regard to the problem size
(i.e., number of state variables and actions), being able to partition it into sub-problems of smaller sizes definitely helps SGPlan to be find a valid
solution quickly. However, there are several reasons that TFD, LPG, and POPF can find overall better quality solutions: (i) their anytime algorithms allow them to
gradually find better quality plans, using the previously found plans as
baseline for pruning unpromising search directions; (ii) SGP's partitioning
algorithm is based on logical relationship between state variables and actions
and ignores all temporal aspects. Thus, combining plans for sub-problems using
logical global constraints can lead to plans of lower quality for
time-sensitive objective function such as minimizing the plan makespan.

\begin{figure*}[t]
  \centering
  \includegraphics[width=0.92\columnwidth]{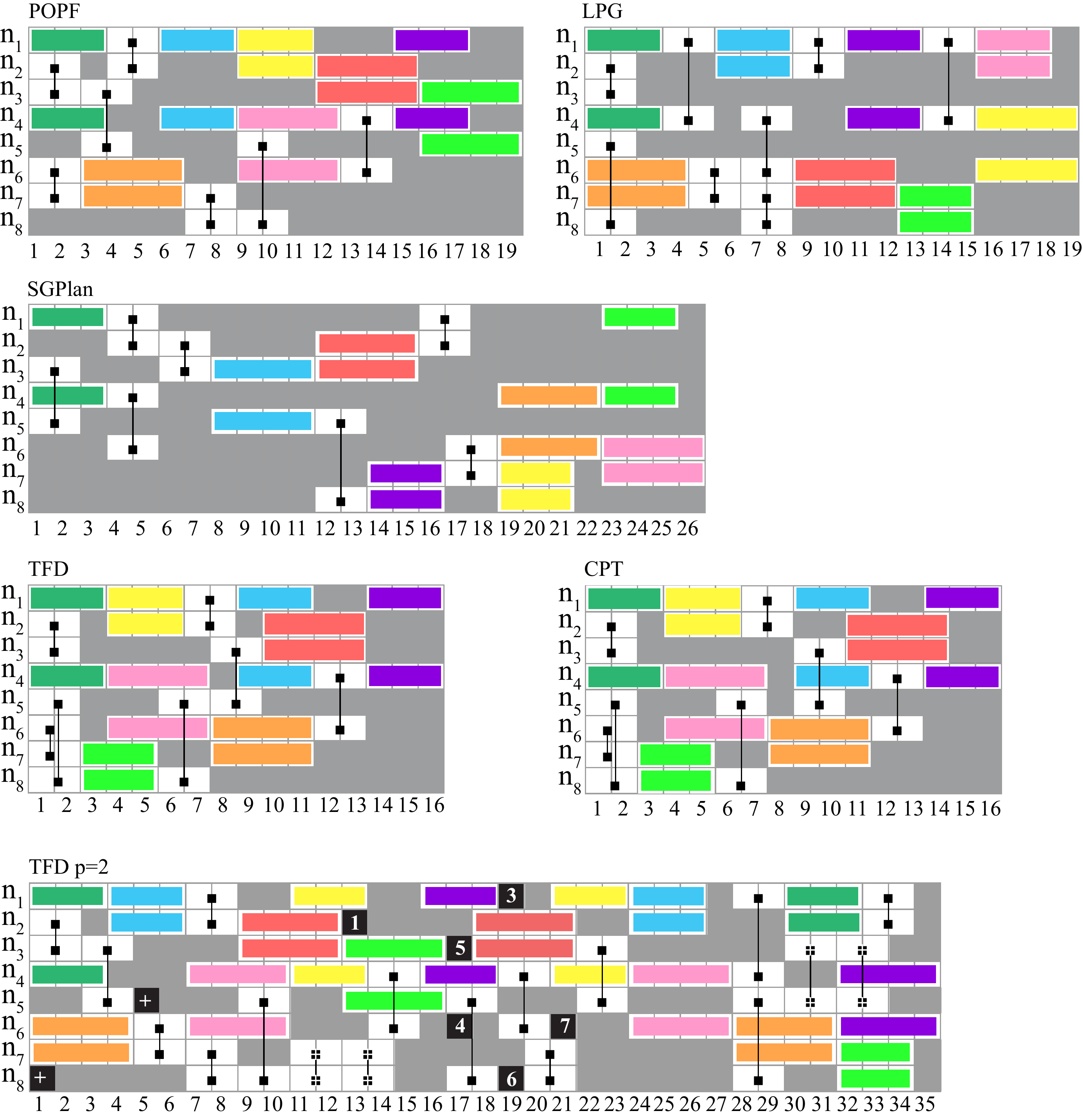}
  \caption{ 
  The Gantt charts show compilations, labeled for each planner of the QAOA for the MaxCut instance depicted in Fig.~\ref{fig:maxcut} on the N=8 processor in
  Fig.~\ref{fig:hardware}. Schedules have time on the x-axis and qubit locations on the y-axis. Each row indicates what gate operates on each qubit at a given time during the plan (Colored blocks represents p-s gates, of duration 3 or 4 depending on their location, with synchronized pair colors associated to edges in Fig.~\ref{fig:maxcut} and white blocks are swap gates).
  The last schedule shows a compilation of $p$ = 2 performed by TFD. Black blocks with numbers are mix gates acting on the corresponding state. Gates marked with a $+$ indicate superfluous gates that were inserted in the plan by TFD, that could be detected and eliminated in post-processing.
}

    \label{fig:gantt}
\end{figure*}

Fig.~\ref{fig:gantt} shows a visualization of plans in a `Gantt chart' format, putting the planners in comparison for a single $N$=8 instance from Fig.~\ref{fig:maxcut}.
To illustrate the representation, we can look at the shown p=2 case: consider the, qstate $q_1$ initially located at $n_1$. The first gate it is involved in is the phase separation gate shown in green between qstates $q1$ and $q4$. The second gate is a phase separation gate between qubits $1$ and $2$ which contain qstates $q1$ and $q3$ respectively, because states $q2$ and $q3$ were swapped in the previous step. State $q1$ is then swapped with the state in qubit $2$, prior to being involved in another phase separation gate, between the contents of qubits $2$ and $3$, this time with state $q5$ that was swapped into qubit $3$ during time steps $3-4$. It is then mixed while still located at qubit $2$. Continuing to read through the chart in this way, we see that qstate $q1$ undergoes the following sequence of actions:
\begin{eqnarray}
\textsc{p-s}_3(q_1,q_4)\rightarrow\textsc{p-s}_3(q_1,q_3)\rightarrow\textsc{p-s}_4(q_1,q_5)\rightarrow\textsc{mix}(q_1)&&\nonumber\\
\rightarrow\textsc{wait}(4)\rightarrow
\textsc{p-s}_4(q_1,q_5)\rightarrow\textsc{wait}(2)&&\nonumber\\
\rightarrow\textsc{p-s}_3(q_1,q3)\rightarrow\textsc{wait}(3)\rightarrow\textsc{p-s}_3(q_1,q_4)&&\nonumber
\end{eqnarray}
where we denote the duration of the $\textsc{p-s}$ gates in subscript and we introduced an $\textsc{wait}$ gate to indicate inaction times. The second mixing phase is trivially scheduled at the end of the last tasks for each qstate.

In the shown case, TFD has found an optimal solution (same makespan as CPT). 

Based on an ``eye-test'' and manual analysis, the best plans returned are usually of good quality but not without defects. Note also that the plan found by TFD for p=2 is also worse than the one that would be trivially obtained by replicating the optimal makespan $p$=1 solution twice (the second time in reverse).
The displayed output also contains some unnecessary gates. Examples are the repeated swaps at time 11 and 30, and the mixing of the un-utilized logical states $q_2$ and $q_8$ at times 1,5. These spurious gates/actions do not affect the makespan, and they can be identified and eliminated by known plan post-processing techniques~\cite{minhdo:icaps03}. We also believe a tighter PDDL model will help eliminate extra gates.

\section{Conclusion and Future Work}
\label{sec:conclusion}

In this paper we presented a novel approach to the problem of compiling
idealized quantum circuits to specific quantum hardware, focusing our
experiments on QAOA circuits. Our presentation and tests have been
focused on the pedagogical and practically relevant example of MaxCut, but the
approach is sufficiently general to be applied to QAOA circuits for
any discrete optimization problem,
and to arbitrary quantum circuits more generally.
Because QAOA has so many commuting gates, we expect to obtain a bigger win for this sort of algorithm than typical quantum algorithms. For most circuits, however, the problem of determining which swaps to make when to ensure that the qstates are next each other in order to carry out the desired gate sequence is highly non-trivial. Many swaps can be done in parallel, and ideally would place qstates in such a way that multiple gates can be carried out before having to swap again, or more generally in a way so as to minimize the run time of the circuit. For these reasons, we expect the temporal planning approach to enable significant gains over a brute force for circuits generally.

A handful of well-established temporal planners were able to compile the
QAOA circuits with reasonable efficiency, demonstrating the viability of this
approach. We plan to expand the portfolio of plannes we have tested, keeping up with latest AI planning technology; the data used in our tests, as well as the PDDL models, will be made available
online at ~\cite{onlinelink}.

One thing that will be expanded in our analysis is the assessment of the quality of the best plans found compared to optimal solutions. At the moment, there is no published work on finding optimal solution for this problem and, as outlined in the previous section, our current effort to get existing optimal-makespan planners to find solutions has been successful only for $N$ = 8, $p$ = 1. 
Another important direction is to support circuit optimization beyond the makespan objective: for instance there is currently tremendous interest in the definition of figure-of-merits that could quantify the power of near-term quantum devices to execute experiment of interests (i.e. quantum supremacy experiments)~\cite{onlinelink2qvolume}.

This work paves the way for future work on the use of
AI planning methods for quantum circuit compilation and design.
In future work, we plan
to further tune the performance of the planners, including choosing an initial
assignment of qstates to qubits favorable for compilation, instead of using a random assignment.  In order to scale
reliably to QAOA circuits with more levels and therefore larger plan sizes, we will develop decomposition approaches in which
$p$ > 1 could be divided into multiple
$p$ = 1 problems to be solved independently and matched in a postprocessing phase. Initial results show that the advantage of not decomposing the problem amounts to approximately 10\% of reduction of the makespan on average, and more than half of the test instances can be scheduled optimally by focusing on the $p$ = 1 problem.

We will also compare with other approaches to this compilation problem such as
sorting networks \cite{Beals13} \cite{brierley2015efficient}
\cite{bremner2016achieving}, constraint programming (CP), or tailored scheduling heuristics~\cite{guerreschi2017gate} and develop more advanced hybrid compilation methods
building on the various strengths of the temporal planning approaches
and of other approaches. Once mature we will integrate compilation with other software supporting experimentation with various near-term processors.
A virtue of the planning approach is that the temporal planning framework is
very flexible with respect to features of the hardware, including irregular
graph structures and diverse gate durations.

In the future, we can include in the PDDL modeling additional features
that are characteristics of quantum computer architectures, such as the
crosstalk effects of 2-qubit gates, realistic durations from optimal 2-qubit gate synthesis that could allow $\textsc{p-s}$ and $\textsc{swap}$ gates to be performed as a single gate, or the ability to \emph{quantum teleport}
quantum states across the chip~\cite{copsey2003toward}, and
features of broad classes of quantum algorithms including
measurement and feedback, error correction, and fault tolerant gate
implementations.
As hardware graphs, primitive gate sets, and gate durations for processors build by experimental
groups become available, we will apply this temporal planning approach
with these hardware parameters as input. Conversely, results from future work comparing results of compilation to different potential architectures may suggest hardware designs that take into account the ability to support efficient compiled circuits.

We will also consider other families of quantum circuits, including QAOA circuits applied to problems beyond MaxCut~\cite{QAOAstuart}, and to more typical quantum circuits, with fewer pairwise commuting gates, but for which it remains difficult to fine an optimal sequence of swap and goal gates. This temporal planning approach to quantum circuit compilation
should be of great interest to the
community developing low-level quantum compilers for generic
architectures \cite{steiger2016projectq} \cite{haner2016software} and to
designers of machine-instructions languages for quantum computing
\cite{smith2016practical} \cite{cross2017open}.\\

\section{Acknowledgements}

The authors acknowledge useful discussions with Will Zeng, Robert
Smith, and Bryan O'Gorman. The authors appreciate support from the NASA
Advanced Exploration Systems program and NASA
Ames Research Center (Sponsor Award No. NNX12AK33A and contract No. NNA16BD14C). The views and conclusions
contained herein are those of the authors and should
not be interpreted as necessarily representing the official policies or endorsements, either expressed or implied,
of the U.S. Government. The U.S.
Government is authorized to reproduce and distribute
reprints for Governmental purpose notwithstanding any
copyright annotation thereon.\\

\bibliographystyle{apalike}
\bibliography{QST}

\end{document}